\documentclass{elsart}
\usepackage{graphics,natbib,graphicx,psfig,amssymb,ccaption} 
\usepackage{lscape} 
\journal{New Astronomy}
\input psfig.sty
\begin{document}
\begin{frontmatter}
\title{A New Absolute Magnitude Calibration for Red Clump Stars}
\author[istanbul]{S. Bilir\corauthref{cor}},
\corauth[cor]{Corresponding author. Fax: +90 212 440 03 70}
\ead{sbilir@istanbul.edu.tr}
\author[istanbul]{T. Ak},
\author[istanbul]{S. Ak},
\author[istanbul]{T. Yontan},
\author[sabanci]{Z.~F. Bostanc\i}
\address[istanbul]{\. Istanbul University, Faculty of Science, 
Department 
of Astronomy and Space Sciences, 34119 University, \. Istanbul, 
Turkey}
\address[sabanci]{Sabanc\i  ~University, Faculty of Engineering 
and Natural 
Sciences, Orhanl\i --Tuzla, 34956 \. Istanbul, Turkey}
\begin{abstract}
We present an $M_{V}$ absolute magnitude calibration including the $B-V$ 
colour and $[Fe/H]$ metallicity for the red clump stars in the 
globular and open clusters with a wide range of metallicities:
$M_{V} = 0.627(\pm 0.104)~(B-V)_{0}+0.046(\pm 0.043)~[Fe/H]+0.262(\pm 0.111).$
The calibration equation is valid in the ranges $0.42<(B-V)_{0}<1.20$ mag, 
$-1.55<[Fe/H]<+0.40$ dex and $0.43< M_{V}<1.03$ mag. 
We found that the consistencies in the comparisons of the distances estimated 
from the calibration equation in this study both with the distances obtained 
from trigonometric parallaxes and spectrophotometric analysis demonstrate that 
reliable precise absolute magnitudes for the clump giants can be estimated 
from the calibration formula. 
\end{abstract}
\end{frontmatter}

\section{Introduction}
Clumping of the core-helium-burning stars to the red end of 
the metal-rich counterpart to the horizontal branch is one 
of the outstanding features in the colour-magnitude diagrams 
of globular and open clusters \citep{Girardi00a}. 
Hence, these giant stars are called the red clump 
(RC) stars or clump giants. Moreover, the RC stars occupy a very 
small region in the colour-magnitude diagram of the stars in 
the solar neighbourhood. Absolute magnitudes of the 
RC stars lie between $M_{V}=+0.7$ and $M_{V}=+1$ mag for 
the stars of spectral types G8 III to K2 III, respectively 
\citep{KeenBarn99}. These features of 
the RC stars make them suitable standard candles for distance 
estimation of stellar clusters and nearby galaxies for which 
the RC stars are resolved \citep{Pietr03}. Moreover, the RC 
stars are one of the most reliable distance indicators 
due to the fact that their absolute magnitude calibrations 
are precisely tied to the {\em Hipparcos} distance scale 
\citep{Alves00,Percival03,Groenewegen08,Laney12,Bilir13}.

\citet{vanHelshoecht07} used the Two Micron All Sky Survey 
\citep[2MASS;][]{Skrutskie06} near-infrared data for a sample 
of 24 open clusters in order to investigate how the $K_s$-band absolute 
magnitude of the RC stars depends on age and metallicity. They 
showed that a constant value of $M_{K_s}=-1.57\pm0.05$ mag is 
a reasonable assumption to use in distance determinations of 
clusters with metallicities of $-0.5\leq [Fe/H]\leq 0.4$ dex 
and ages of $0.31\leq t\leq 7.94$ Gyr. This absolute magnitude 
value was also confirmed by \citet{Groenewegen08} who found 
$M_{K_s}=-1.54\pm0.04$ mag using the newly reduced 
{\em Hipparcos} trigonometric data \citep{vanLeeuwen2007}. On 
the other hand, \citet{Bilir13} showed 
that the absolute magnitudes of RC stars depend on two-colour 
indices defined in the Johnson-Cousins ($BVI$), 2MASS ($JHK_{s}$) 
and Sloan Digital Sky Survey (SDSS; $gri$) photometries.

Although it is possible to define mean colours and absolute 
magnitudes for the RC stars in different photometric bands and 
the mean values have been suggested in some studies 
\citep[e.g.][]{Alves00,Groenewegen08,Laney12,Bilir13}, it is well known 
that the RC stars occupy certain ranges of colours and 
absolute magnitudes in the colour-magnitude diagrams. Thus, 
absolute magnitudes of the RC stars depend on the metallicity 
and colours especially in optical bands, although it is found 
that the metallicity dependence is weak \citep{Bilir13}.

The mean $V$ and $I$ magnitudes of the RC 
stars can be considerably affected by stellar population differences. 
\citet{Udalski00} found a weak dependence of the $I$-band absolute magnitude of 
the RC stars in the solar neighbourhood on $[Fe/H]$ at the level of 
$\sim 0.13$ mag dex$^{-1}$ and suggested an absolute magnitude correction of 
0.07 mag in the $I$-band for Large Magellanic Cloud. 
Using the models of \citet{Girardi00b}, \citet{Girardi01}  predicted 
mean RC absolute magnitudes in $V$ and $I$ bands for a large range of ages and 
metallicities. Their models clearly showed that the absolute magnitudes of 
the RC stars in $I$-band depends both on metallicity and age, so that the 
metallicity dependence must be taken into account in the absolute magnitude 
calculations of the RC stars \citep{Salaris02}.

Thus, instead 
of a general mean for the absolute magnitude, an absolute 
magnitude calibration including at least one colour and 
metallicity should be used in the distance estimation of the 
RC stars. As more and more distant objects are observed in the 
sky surveys, metallicity-dependent uncertainties in the 
absolute magnitude determinations of the RC stars increase the 
error levels in the estimation of distances and physical 
parameters of the objects including these stars, such as stellar 
clusters.  

Aim of this study is to derive an $M_{V}$ absolute magnitude 
calibration including the $B-V$ colour and $[Fe/H]$ metallicity 
using the RC stars in the globular and open clusters with a wide 
range of metallicities. In the next section we describe how the 
data for the RC stars were collected from the literature. In the 
Section 3, we derive an absolute magnitude calibration for the RC 
stars and compare the distances of the RC stars derived from our 
calibration formula with those obtained from the {\em Hipparcos} 
trigonometric data and spectrophotometric data. Finally, we 
conclude and discuss the results in the Section 4.

\section{The data}
The data in this study were collected from the colour-magnitude 
diagrams of the globular and open clusters in our Galaxy. The data for 
globular clusters were taken from \cite{Piotto02} who analysed 
{\it Hubble Space Telescope} observations. \cite{Piotto02} examined 74 
globular clusters in {\it F439W} and {\it F555W} bands and analysed 
their colour-magnitude diagrams. They transformed the original 
magnitudes to $BV$ photometry using calibrations given in their 
study. It is clearly seen that 25 of these globular clusters show 
remarkable clumping in the RC region in their colour-magnitude 
diagrams \citep[see Fig. 4 in][]{Piotto02}. 
Borders of the RC regions were determined by eye inspection of the 
colour-magnitude diagrams of globular clusters. Distance moduli, 
colour excesses and metallicities of the globular clusters were 
primarily taken from \cite{Piotto02}. In cases where newer 
measurements exist, we used the new ones. The Galactic coordinates 
($l$, $b$), distance moduli ${(m-M)}_{V}$, colour excesses $E(B-V)$ 
and metallicities $[Fe/H]$ for the globular clusters are listed 
in Table 1. 

$B-V$ colours and $V$ magnitudes of the RC stars were 
de-reddened and their $M_{V}$ absolute magnitudes were calculated 
using distance moduli given in Table 1. 
Frequency distributions of 
the de-reddened colours $(B-V)_{0}$ and absolute magnitudes $M_{V}$ 
of the RC stars in each globular cluster were estimated and central 
positions of the peaks in the distributions obtained for each 
globular cluster were found by fitting Gaussian functions. 
$(B-V)_{0}-V_{0}$ colour-magnitude diagrams and frequency 
distributions of the $(B-V)_{0}$ colours and $M_{V}$ absolute 
magnitudes of the globular clusters listed in Table 1 are 
shown in Fig. 1. Central positions of the peaks and standard 
deviations for the $(B-V)_{0}$, $M_{V}$ absolute magnitude 
distributions and the number of RC stars in 25 globular clusters in 
our sample are given in Table 1. 

As the aim of this study is to derive a colour ($B-V$) and metallicity 
($[Fe/H]$) dependent absolute magnitude ($M_{V}$) calibration of the 
RC stars, metallicity range of the sample must be as large as possible. 
Metallicity range of the globular clusters in our sample is 
$-1.55<[Fe/H]<-0.30$ dex. In order to extend this metallicity range, 
the RC stars in the open clusters must be included in our sample. 
However, in general there are a few RC stars in open clusters. 
In this study, we selected the open clusters, where the RC stars 
prominently detectable in their colour-magnitude diagrams, from the 
studies by \citet{Twarog97,Grocholski02,Percival03,vanHelshoecht07}.
We listed their colour excesses, distance moduli and metallicities 
of these open clusters in Table 1. We determined borders of the RC 
region by eye inspection of the colour-magnitude diagrams of open 
clusters. As the number of RC stars is small, we simply calculated the 
average $(B-V)_{0}$ colour indices and  $M_{V}$ absolute magnitudes 
of these stars (see Table 1). By including the selected open clusters 
in the sample, upper limit of the metallicities were extended to about 
+0.40 dex. The selected open clusters increased the total number of 
the clusters in the sample to 31 (Table 1). 

As the metallicities were collected from the literature that includes 
continually changing average of parameters, the metallicity 
scale used in our study should be carefully taken into account. 
One of the most important metallicity scales was given by \citet[ZW,][]{Zinn84}. 
They used photometric indices of clusters calibrated primarily by Cohen's 
(\citeyear{Cohen83}) spectroscopic data. As the distance moduli, colour 
excesses and metallicities for 18 of 25 globular clusters in Table 1 were 
taken from \citet{Piotto02} and the 2010 version of the Harris catalog 
\citep{Harris96} whose metallicity scales fit to ZW metallicity scale, 
we conclude that the metallicities in our globular cluster data fit to 
this scale, in general. We also checked for the metallicity scale of the 
remaining seven globular clusters in Table 1 that were not taken from 
the Harris catalog \citep{Harris96} and found that these metallicities 
are in a good agreement with the ZW scale.

\begin{table*}
\setlength{\tabcolsep}{4pt}
{\scriptsize
\begin{center}
\caption{The Galactic coordinates ($l$, $b$), $(m-M)_{V}$ distance moduli, 
$E(B-V)$ colour excesses and $[Fe/H]$ metallicities collected from the 
literature for the clusters in this study. Central positions of 
the Gaussian fits applied to the distributions of the de-reddened 
$(B-V)_{0}$ colours and $M_{V}$ absolute magnitudes of the RC stars in 
the globular clusters are given. The $(B-V)_{0}$ colours and $M_{V}$ 
absolute magnitudes of the RC stars in the open clusters are also indicated. 
Number of the RC stars for each cluster are given in the last column.}
\begin{tabular}{clccccccccc}
\hline
ID & Cluster & $l$ & $b$ & $E(B-V)$ & $(m-M)_V$ & $[Fe/H]$ & Refs. & $(B-V)_o$ & $M_V$ &  N \\
   &         & ($^{o}$) & ($^{o}$) & (mag) & (mag) & (dex) &  & (mag) &   (mag)  &  \\
\hline
\multicolumn{11}{l}{Globular Clusters} \\
1  & NGC 6723 &   0.07 & -17.30 & 0.05 & 14.87 & -1.12 & 1   & 0.574 $\pm$ 0.045 & 0.614 $\pm$ 0.067 & ~~58\\
2  & NGC 6569 &   0.48 &  -6.68 & 0.53 & 16.83 & -0.76 & 2   & 0.695 $\pm$ 0.054 & 0.697 $\pm$ 0.065 & ~155\\
3  & NGC 6652 &   1.53 & -11.38 & 0.09 & 15.28 & -0.81 & 2   & 0.777 $\pm$ 0.031 & 0.730 $\pm$ 0.036 & ~~51\\
4  & NGC 6637 &   1.72 & -10.27 & 0.18 & 15.28 & -0.64 & 2   & 0.762 $\pm$ 0.025 & 0.686 $\pm$ 0.079 & ~121\\
5  & NGC 6624 &   2.79 &  -7.91 & 0.28 & 15.36 & -0.44 & 2   & 0.824 $\pm$ 0.051 & 0.755 $\pm$ 0.047 & ~114\\
6  & NGC 6171 &   3.37 &  23.01 & 0.33 & 15.05 & -1.02 & 2   & 0.775 $\pm$ 0.014 & 0.625 $\pm$ 0.050 & ~~26\\
7  & NGC 6356 &   6.72 &  10.22 & 0.28 & 16.77 & -0.50 & 1   & 0.826 $\pm$ 0.038 & 0.761 $\pm$ 0.069 & ~328\\
8  & NGC 6440 &   7.73 &   3.80 & 1.07 & 17.95 & -0.34 & 1   & 1.025 $\pm$ 0.043 & 0.891 $\pm$ 0.103 & ~466\\
9  & NGC 6638 &   7.90 &  -7.15 & 0.43 & 16.40 & -1.00 & 3   & 0.569 $\pm$ 0.033 & 0.494 $\pm$ 0.025 & ~63\\
10 & NGC 6864 &  20.30 & -25.75 & 0.16 & 17.09 & -1.29 & 2   & 0.607 $\pm$ 0.048 & 0.625 $\pm$ 0.042 & ~182\\
11 & NGC 6539 &  20.80 &   6.78 & 0.97 & 17.63 & -0.66 & 1   & 0.887 $\pm$ 0.047 & 0.780 $\pm$ 0.115 & ~123\\
12 & NGC 6712 &  25.35 &  -4.32 & 0.45 & 15.60 & -1.01 & 1   & 0.707 $\pm$ 0.046 & 0.639 $\pm$ 0.062 & ~~46\\
13 & NGC 6934 &  52.10 & -18.89 & 0.09 & 16.48 & -1.54 & 1   & 0.425 $\pm$ 0.012 & 0.444 $\pm$ 0.016 & ~~41\\
14 & NGC 1851 & 244.51 & -35.04 & 0.02 & 15.58 & -1.31 & 4, 5& 0.609 $\pm$ 0.031 & 0.561 $\pm$ 0.043 & ~126\\
15 & NGC 1261 & 270.54 & -52.12 & 0.01 & 16.10 & -1.27 & 5   & 0.605 $\pm$ 0.020 & 0.629 $\pm$ 0.073 & ~~57\\
16 & NGC 2808 & 282.19 & -11.25 & 0.18 & 15.61 & -0.93 & 6   & 0.645 $\pm$ 0.030 & 0.652 $\pm$ 0.050 & ~340\\
17 & NGC ~~362& 301.53 & -46.25 & 0.04 & 14.94 & -1.16 & 7   & 0.537 $\pm$ 0.037 & 0.559 $\pm$ 0.020 & ~157\\
18 & NGC ~~104& 305.90 & -44.89 & 0.04 & 13.37 & -0.76 & 1   & 0.741 $\pm$ 0.033 & 0.658 $\pm$ 0.023 & ~283\\
19 & NGC 6362 & 325.55 & -17.57 & 0.09 & 14.68 & -0.99 & 2   & 0.582 $\pm$ 0.029 & 0.638 $\pm$ 0.022 & ~~24\\
20 & NGC 5927 & 326.60 &   4.86 & 0.45 & 15.82 & -0.49 & 2   & 0.932 $\pm$ 0.034 & 0.870 $\pm$ 0.056 & ~179\\
21 & NGC 6584 & 342.14 & -16.41 & 0.10 & 15.95 & -1.49 & 1   & 0.525 $\pm$ 0.052 & 0.525 $\pm$ 0.062 & ~~39\\
22 & NGC 6388 & 345.56 &  -6.74 & 0.40 & 16.54 & -0.60 & 1   & 0.791 $\pm$ 0.059 & 0.672 $\pm$ 0.068 & 1038\\
23 & NGC 6441 & 353.53 &  -5.01 & 0.51 & 17.04 & -0.53 & 8   & 0.884 $\pm$ 0.038 & 0.781 $\pm$ 0.093 & 1197\\
24 & NGC 6304 & 355.83 &   5.38 & 0.52 & 15.54 & -0.59 & 1   & 0.875 $\pm$ 0.015 & 0.764 $\pm$ 0.066 & ~~87\\
25 & NGC 6316 & 357.18 &   5.76 & 0.51 & 17.08 & -0.55 & 9   & 0.879 $\pm$ 0.053 & 0.796 $\pm$ 0.039 & ~177\\
\multicolumn{11}{l}{Open Clusters} \\
26 & NGC 6791 &  69.96 &  10.90 & 0.16 & 13.56 &  0.37 & 10, 11 & 1.197 $\pm$ 0.025 & 1.020 $\pm$ 0.030 & ~~23\\
27 & NGC 7789 & 115.53 &  -5.38 & 0.29 & 12.12 & -0.13 & 12, 13 & 0.928 $\pm$ 0.035 & 0.800 $\pm$ 0.095 & ~~24\\
28 & NGC 188  & 122.86 &  22.38 & 0.09 & 11.45 & -0.03 & 12, 14 & 1.110 $\pm$ 0.030 & 0.988 $\pm$ 0.061 & ~~~3\\
29 & Be 39    & 223.46 &  10.09 & 0.11 & 13.31 & -0.15 & 12, 15 & 1.087 $\pm$ 0.040 & 0.954 $\pm$ 0.090 & ~~~6\\
30 & NGC 2477 & 253.56 &  -5.84 & 0.23 & 11.45 &  0.00 & 12, 15 & 0.990 $\pm$ 0.055 & 0.915 $\pm$ 0.098 & ~~52\\
31 & Mel 66   & 259.56 & -14.24 & 0.14 & 13.65 & -0.38 & 12, 16 & 0.962 $\pm$ 0.010 & 0.846 $\pm$ 0.070 & ~~15\\
\hline
\end{tabular} 
\end{center}
1) \citet{Piotto02}, 2) \citet{Harris10}, 3) \citet{Valenti07}, 4) \citet{Walker98}, 
5) \citet{Gratton10}, 6) \citet{Milone12}, 7) \citet{Szekely07}, 8) \citet{Matsunaga09}, 
9) \citet{Layden03}, 10) \citet{Sandage03}, 11) \citet{Stetson03}, 
12) \citet{Percival03}, 13) \citet{Girardi00a}, 14) \citet{Sarajedini99}, 
15) \citet{Kassis97}, 16) \citet{Twarog95}.\\    
}
\end{table*}

Dependence of the mean $M_{V}$ absolute magnitude of the 
RC stars in the sample clusters both on $[Fe/H]$ metallicity and 
$(B-V)_{0}$ colour are presented in Fig. 2a-b. It is clearly seen that 
the RC stars in our study comprise a considerably wide range of 
$(B-V)_{0}$ colours, $0.4\leq (B-V)_{0}\leq 1.2$ mag. Metallicity 
of the clusters covers a wide range as well, $-1.55<[Fe/H]<+0.40$ dex. 
$M_{V}$ absolute magnitudes of the RC stars in our sample range from 
0.43 to 1.03 mag. Metal-poor halo clusters are found in the bluer 
$(B-V)_{0}$ colours in Fig. 2 while metal-rich bulge clusters 
are redder ($0.6\leq (B-V)_{0}\leq 0.9$ mag). In our sample, $B-V$ 
colours of open clusters in the thin disc are larger than 0.9 mag.

There is a linear relation with a slope of 2.97 dex mag$^{-1}$ between the 
$[Fe/H]$ metallicity and $M_{V}$ absolute magnitude of the RC stars in Fig. 2a. 
The scatter in this relation ($\sim$0.1 mag) is larger than the 
scatter ($\sim$0.04 mag) in the relation between the $M_V$ absolute 
magnitude and $(B-V)_{0}$ colour of the RC stars. Fig. 2c shows the linear 
relation between the $(B-V)_{0}$ colour indices and 
$[Fe/H]$ metallicities for the clusters in our sample. Slope of the relation
is 2.44 dex mag$^{-1}$. Fig. 2a-c demonstrate that the population effects on 
the absolute magnitudes and colour indices of the RC stars can not be negligible 
in the Galactic scales.

\section{The absolute magnitude calibration}

From the considerations above, it is clear that the $M_{V}$ absolute 
magnitude of the RC stars is well correlated both with the $[Fe/H]$ 
metallicity and $(B-V)_{0}$ colour. Thus, a regression analysis based 
on 31 clusters in Table 1 gives the following relation:

\begin{equation}
M_{V} = 0.627(\pm 0.104)~(B-V)_{0} + 0.046(\pm 0.043)~[Fe/H] + 0.262(\pm 0.111).
\end{equation}

Numbers in parenthesis are standard errors of the coefficients. The correlation 
coefficient and standard deviation of the calibration are $R=0.971$ and 
$s=0.036$ mag, respectively. T-scores of the variables $(B-V)_{0}$, $[Fe/H]$ and the 
constant in Eq. 1 are 6.01 ($p = 0.00$), 1.06 ($p = 0.30$) and 2.35 ($p = 0.03$), respectively, 
with a degree of freedom of 29. This equation is reliable and valid in the ranges 
$0.42<(B-V)_{0}<1.20$ mag, $-1.55<[Fe/H]<+0.40$ dex and $0.43<M_{V}<1.03$ mag. 

A comparison of the absolute magnitudes calculated from the calibration with 
the original absolute magnitudes in Table 1 is shown in Fig. 3. This figure 
demonstrates that 22 of the clusters ($\sim$ 71$\%$) are within 1$\sigma$ 
deviation. There is no any systematical tendency in the lower panel of Fig. 3 
that shows residuals from the calibration equation. 

T-score analysis shows that metallicity dependence of the calibration is relatively weak, 
while the dependence on the colour $(B-V)_{0}$ is strong. If the metallicity is not 
taken into account, the absolute magnitude changes $\sim$0.1 mag for the stars with 
lowest and highest metallicities. It should be also noted that the absolute magnitude 
decreases with decreasing metallicity (see Fig. 2). That is why we emphasize that 
the metallicity, if it is known, should be taken into account in the absolute magnitude 
calibrations of the RC stars. Although the metallicity dependence of the Eq. 1 is found 
weak, we suggest that absolute magnitude estimation errors related with the metallicity 
will be more important for the distant RC stars.

\subsection{The application}

Using the absolute magnitude calibration in this study, $M_{V}$ absolute 
magnitudes were estimated for the RC stars collected from two studies, 
i.e. \cite{Laney12} and \cite{Saguner11}. The RC stars with $[Fe/H]$ 
metallicities are listed in \cite{Laney12} who collected metallicities 
from \cite{Liu07} and \cite{McWilliam90}. They observed 226 RC stars in the  
near-infrared $JHK$-bands in the Solar neighbourhood and calculated 
mean absolute magnitudes of the RC stars using {\em Hipparcos} 
parallaxes \citep{vanLeeuwen2007}. In our study, $B-V$ colours were 
taken from newly reduced {\em Hipparcos} catalogue \citep{vanLeeuwen2007}, 
parallaxes and $[Fe/H]$ metallicities from \cite{Laney12} who collected 
$[Fe/H]$ metallicities of only 101 RC stars from \cite{Liu07} and 
\cite{McWilliam90}. $[Fe/H]$ metallicities of these stars are between 
-0.6 and 0.4 dex. As \cite{Laney12} scaled the metallicities taken 
from \cite{McWilliam90} according to the ones from \cite{Liu07}, it can 
be concluded that the metallicities in Laney's study are in the same 
scale and self-consistent.

We first de-reddened the B-V colour indices using the $E(B-V)$ colour-excesses 
evaluated for the 101 program stars using the maps of \citet*{Schlegel98}, 
and this was reduced to a value corresponding to the distance of the star by 
means of the equations of \citet{Bahcall80}. For a detailed explanation of the 
de-reddening procedure, see \citet{Bilir13}. The reduced $E(B-V)$ colour-excesses 
of the program stars are between 0.001 and 0.081 mag with a median value of 0.010 mag.
We calculated $M_{V}$ absolute magnitudes of these 101 RC stars using the 
absolute magnitude calibration derived in this study. 

We estimated distances putting our calibration results into the Pogson 
equation. Fig. 4 shows the  comparison of distances from our calibration with 
those taken from \cite{Laney12}. We note that \cite{Laney12} selected the 
distances of the RC stars in their study from {\em Hipparcos} catalogue
\citep{vanLeeuwen2007}. The distances calculated from our calibration are 
in very well agreement with those taken from {\em Hipparcos} 
catalogue \citep{Laney12}. The mean and standard deviation of the 
residuals are only -3.5 and 6.2 pc, respectively. A considerable part of 
the sample is in 1$\sigma$ limits in Fig. 4. This comparison demonstrates that 
the distances based on the $M_{V}$ absolute magnitude calibration of the RC 
stars in the stellar clusters are as precise as comparable to the distances 
derived from {\em Hipparcos} parallaxes of the RC stars within 150 pc.

We also compared the distances estimated via our calibration with 
those obtained by a spectrophotometric method. \cite{Saguner11} 
calculated spectrophotometric distances of 305 faint and 
high Galactic latitude RC stars by deriving their atmospheric 
parameters from high signal-to-noise middle resolution spectra. We 
removed 54 of 305 stars from the sample as they are out of the validity 
limits of the absolute magnitude calibration in this study, reducing 
the number of stars to 251. While the $[Fe/H]$ metallicities and $A_{V}$ total 
absorptions were taken from \cite{Saguner11}, $V$ apparent 
magnitudes and $B-V$ colours from {\em Hipparcos} catalogue 
\citep{vanLeeuwen2007}. The $[Fe/H]$ metallicities are between 
-0.54 and 0.42 dex. Since the metallicities were found by \cite{Saguner11}, 
we conclude that the metallicities in this study are in the same scale 
and self-consistent.

After apparent magnitudes and colours 
were de-reddened, absolute magnitudes and distances of 251 RC stars 
listed in \cite{Saguner11} were estimated using the absolute 
magnitude calibration derived in this study. In Fig. 5, we compared 
the distances of the RC stars estimated using our absolute magnitude 
calibration with the spectrophotometric distances taken from 
\cite{Saguner11} in which the RC stars lie between 200 and 
500 pc from the Sun. Note that the distances taken from \cite{Laney12} 
are smaller than 150 pc. Fig. 5 demonstrates that the distances 
obtained from two different methods are very well correlated with 
a scatter of $s=11$ pc. As the mean distance difference is only 
$<\Delta d>=-0.4$ pc (see Fig. 5b), we conclude that there is no any 
systematic difference between these distances.

\section{Conclusions}

We derived an $M_{V}$ absolute magnitude calibration in terms of the $B-V$ 
colour and $[Fe/H]$ metallicity using the RC stars in the globular and 
open clusters with a wide range of metallicities. The comparisons of the 
distances estimated from the calibration equation in this study both with 
the distances obtained from trigonometric parallaxes and spectrophotometric 
analysis indicate that the calibration formula gives very reliable absolute 
magnitudes for the estimation of distances for the RC stars up to 500 pc. 
We conclude in this study that, instead of using a unique absolute magnitude 
for all the RC stars, an absolute magnitude calibration in the optical bands 
based on at least one colour and metallicity should be used in the precise 
distance estimates.

Taking into account the metallicity in the calculation of absolute magnitudes 
of the RC stars can be important in the shallow/deep photometric and 
spectroscopic surveys, such as the RAdial Velocity Experiment (RAVE) and SDSS, 
and in the distance calculations of the stellar clusters or field RC stars. 
With the help of the absolute magnitude calibration derived in this study, 
and with the uncertainties in the metallicity dependence of the Cepheid 
period-luminosity relation \citep{Percival03,Majaess12}, the RC stars can 
serve as standard candles for the distance calculations of nearby 
Galaxies, as well. This powerful calibration can be also used in the 
derivation of the Galactic model parameters in the optical bands.

\section{Acknowledgments}

We are grateful to Prof. Dr. Salih Karaali for the inspiration and helpful discussions.   
We would like to thank the referee for comments and suggestions. We also 
thank to Dr. Tolga G\"uver for checking the manuscript. This study was supported 
in part by the Scientific and Technological Research Council (T\"UB\.ITAK) under 
project numbers 112T120 and 111T650. This research has made use of NASA's 
Astrophysics Data System and the SIMBAD database, operated 
at CDS, Strasbourg, France.

\pagebreak[4]

\begin{figure}
\begin{center}
\includegraphics[scale=.9, angle=0]{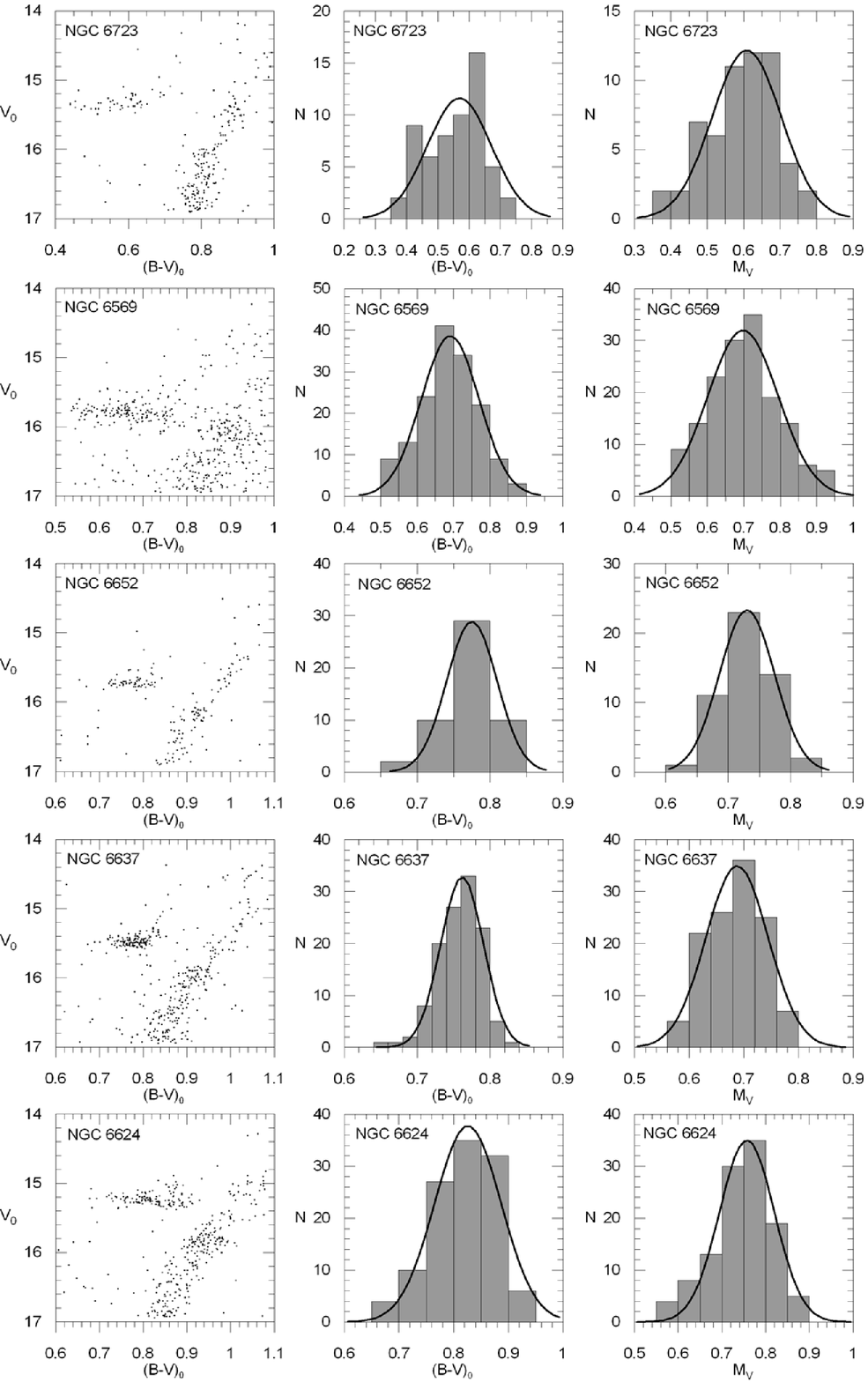}
\caption[]{Colour-magnitude diagrams and distributions of 
the $(B-V)_{0}$ colours and $M_{V}$ absolute magnitudes of the globular 
clusters listed in Table 1.}
\end{center}
\end{figure}

\begin{figure}
\contcaption{} 
\begin{center}
\includegraphics[scale=0.9, angle=0]{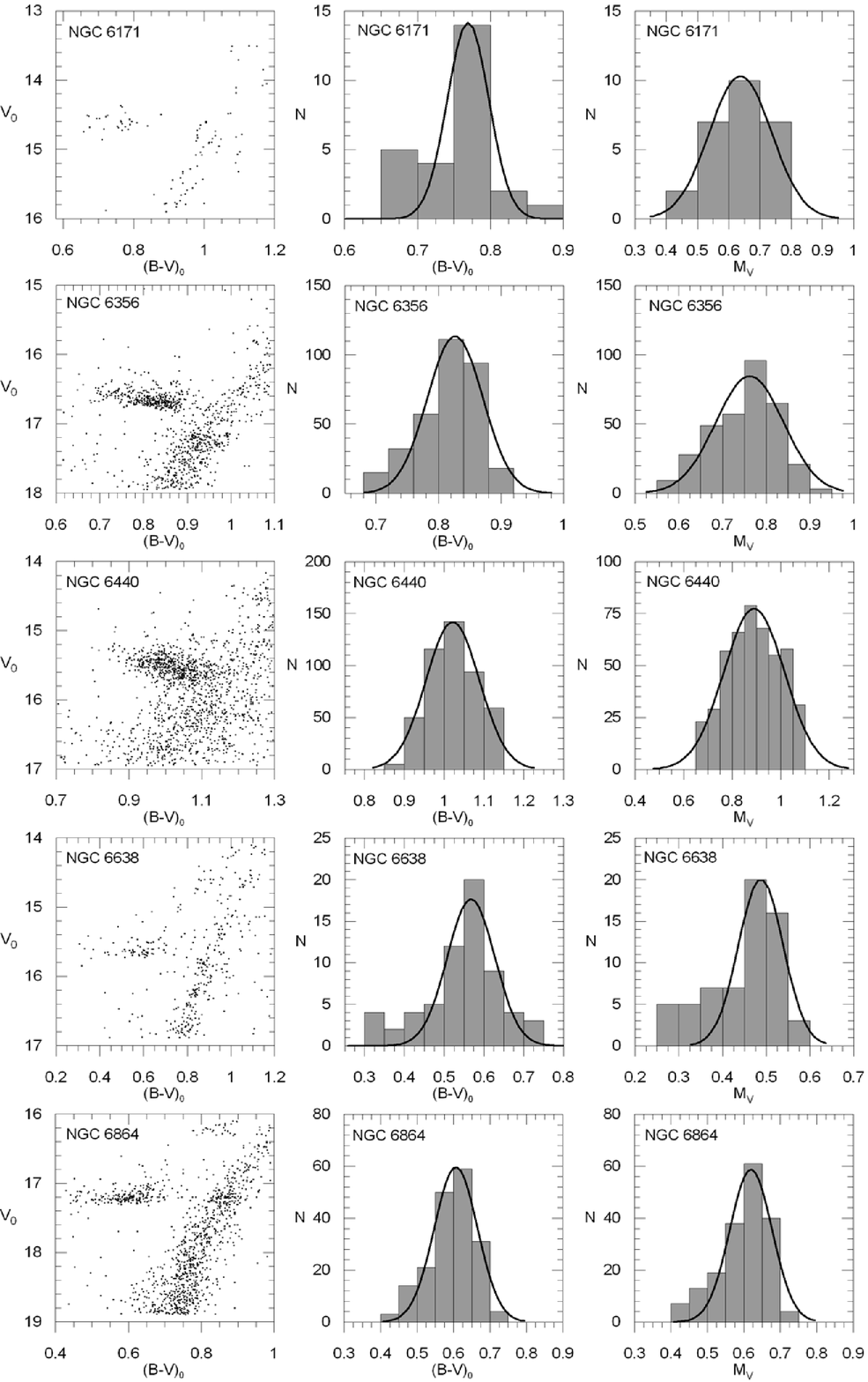}
\end{center}
\end{figure}

\begin{figure}
\contcaption{} 
\begin{center}
\includegraphics[scale=0.9, angle=0]{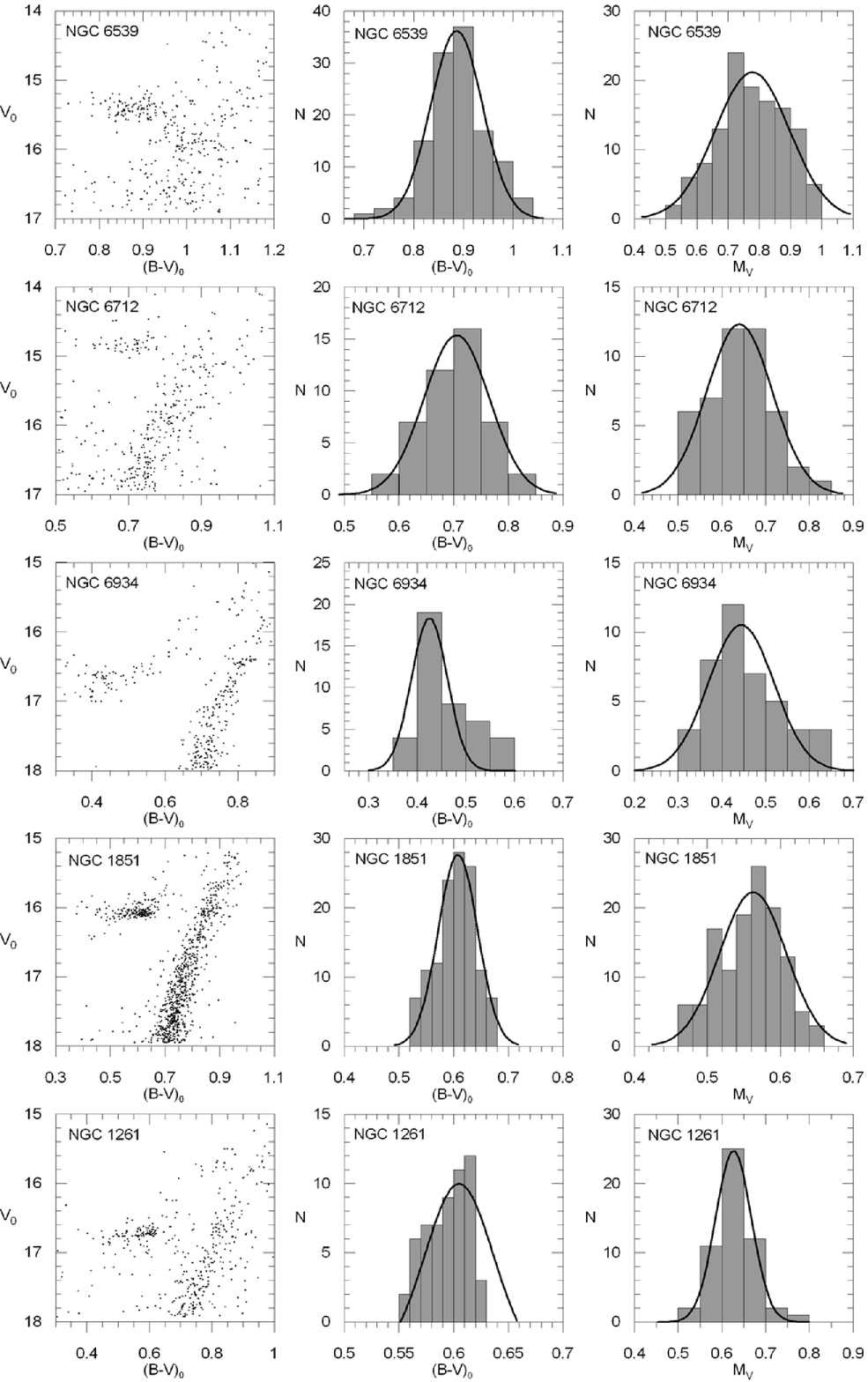}
\end{center}
\end{figure}

\begin{figure}
\contcaption{} 
\begin{center}
\includegraphics[scale=0.9, angle=0]{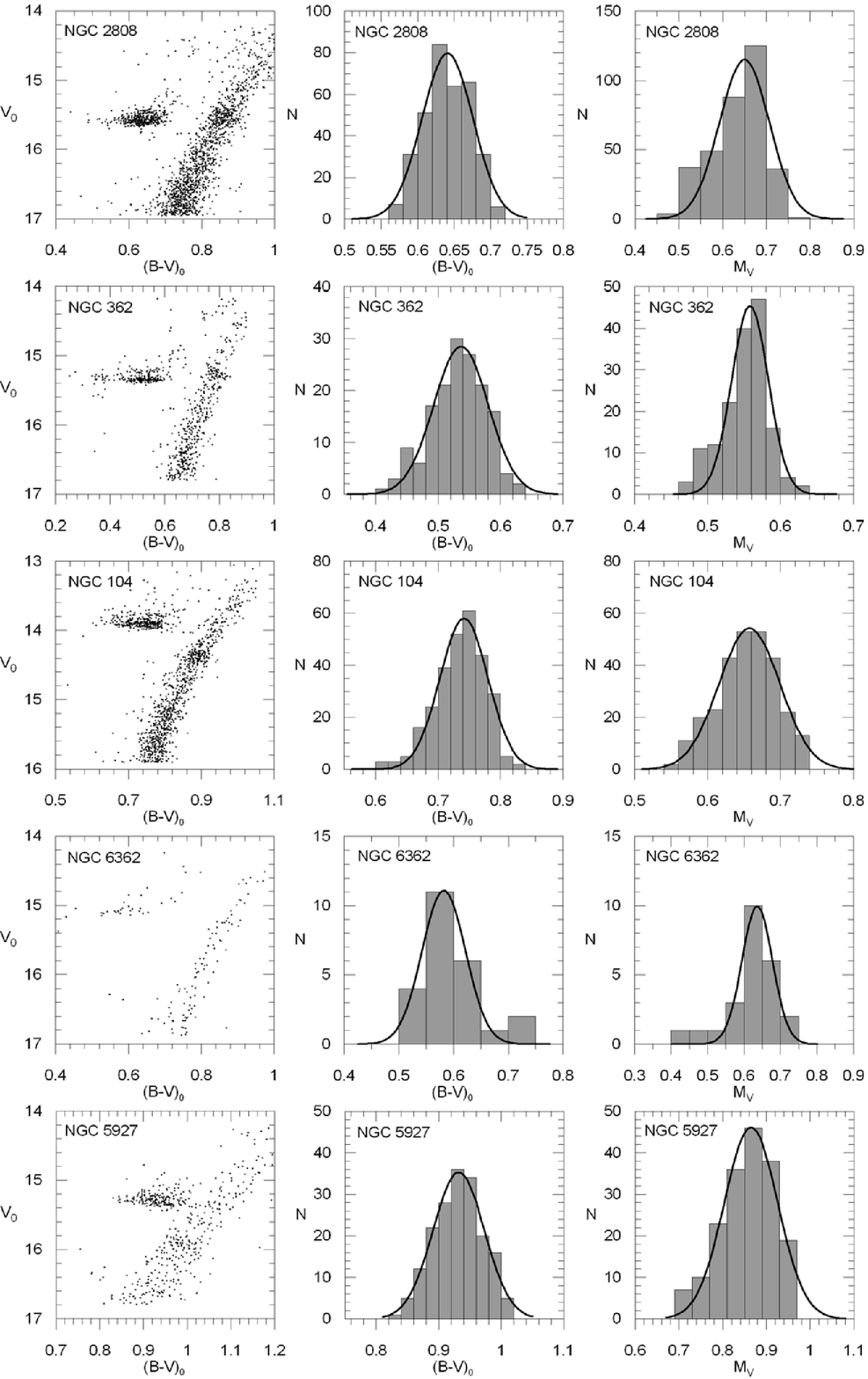}
\end{center}
\end{figure}

\begin{figure}
\contcaption{} 
\begin{center}
\includegraphics[scale=0.9, angle=0]{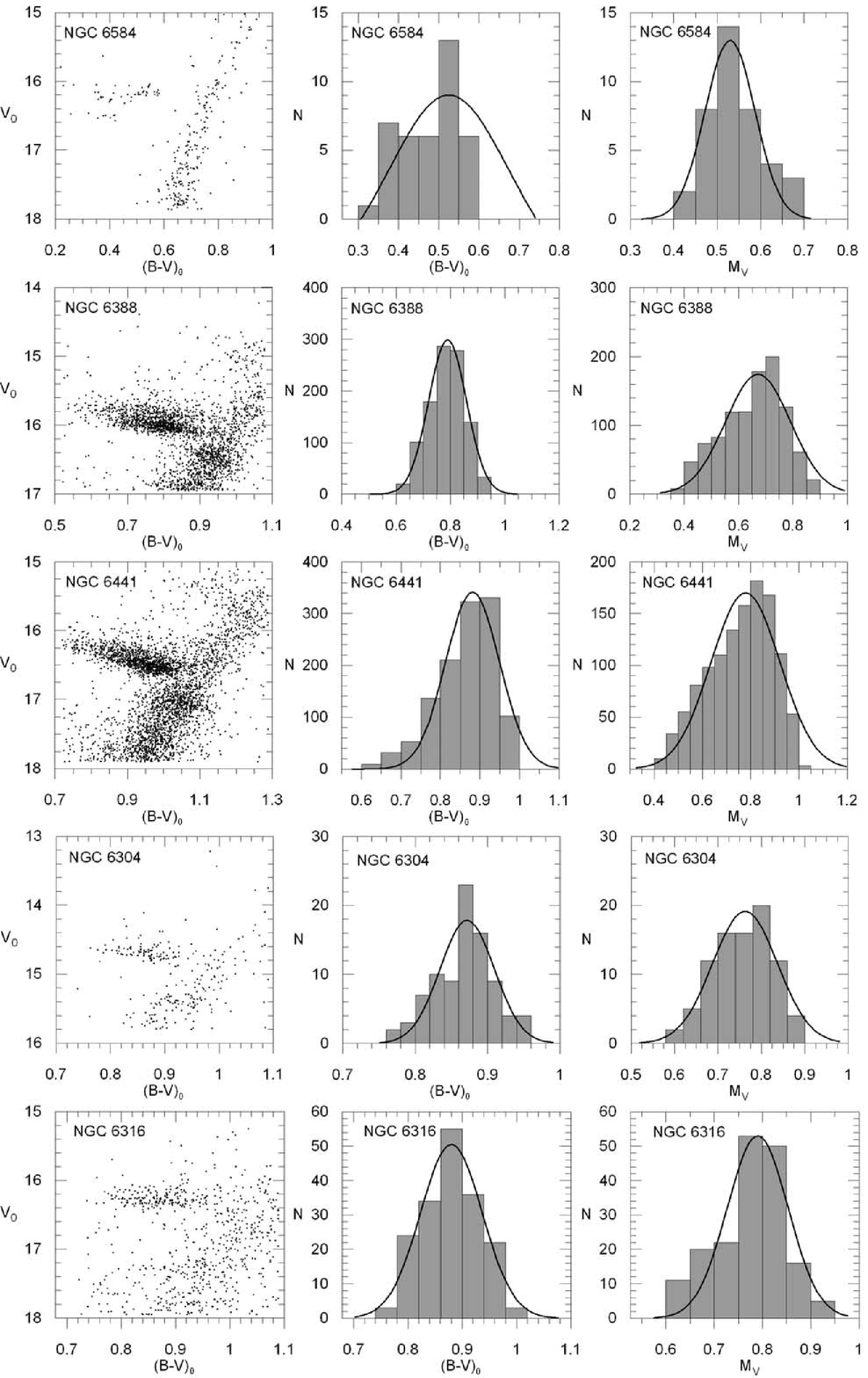}
\end{center}
\end{figure}

\begin{figure}
\begin{center}
\includegraphics[scale=0.6, angle=0]{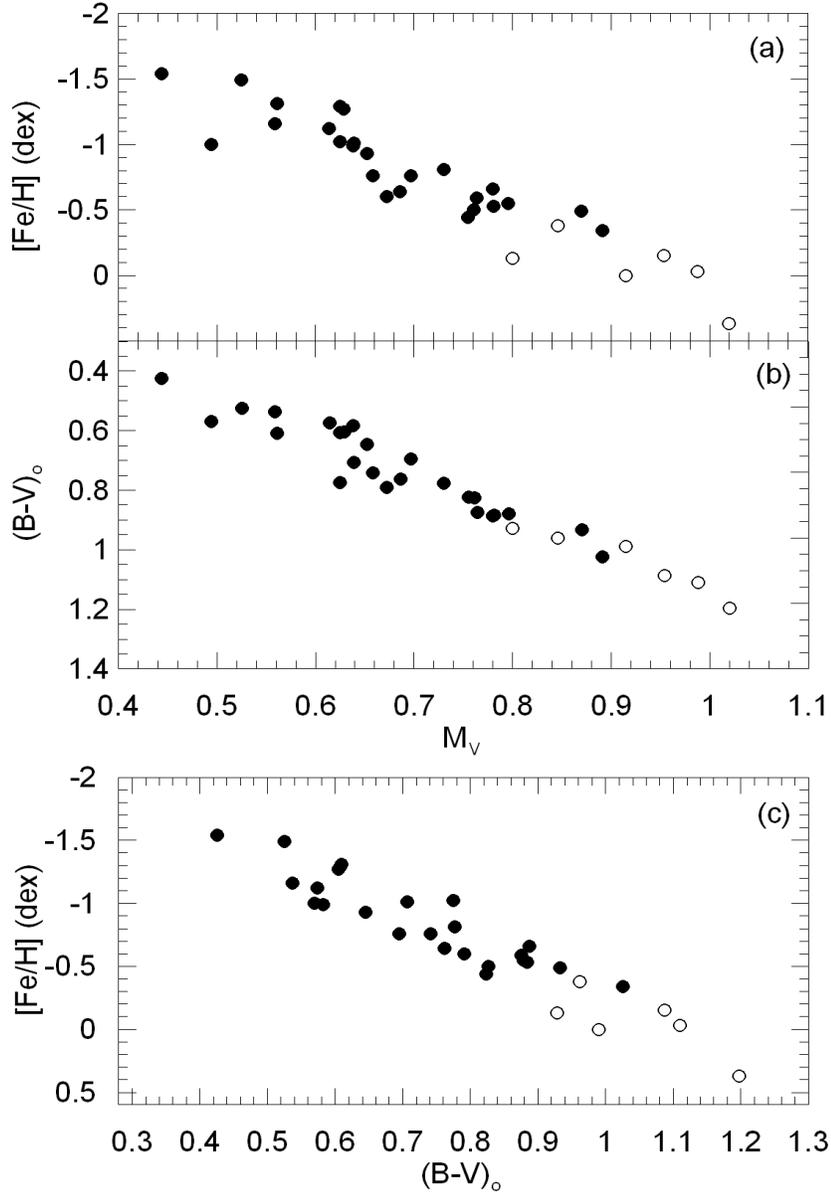}
\caption[]{Dependence of the mean absolute magnitude of the RC 
stars in the sample clusters on $[Fe/H]$ metallicity (a) and 
$(B-V)_{0}$ colour (b). Relation between the $(B-V)_{0}$ colours 
and $[Fe/H]$ metallicities of the RC stars is shown in the panel c. 
Filled and open circles denote the data from globular and open 
clusters, respectively. The data are taken from Table 1.}
\end{center}
\end{figure}

\begin{figure}
\begin{center}
\includegraphics[scale=0.6, angle=0]{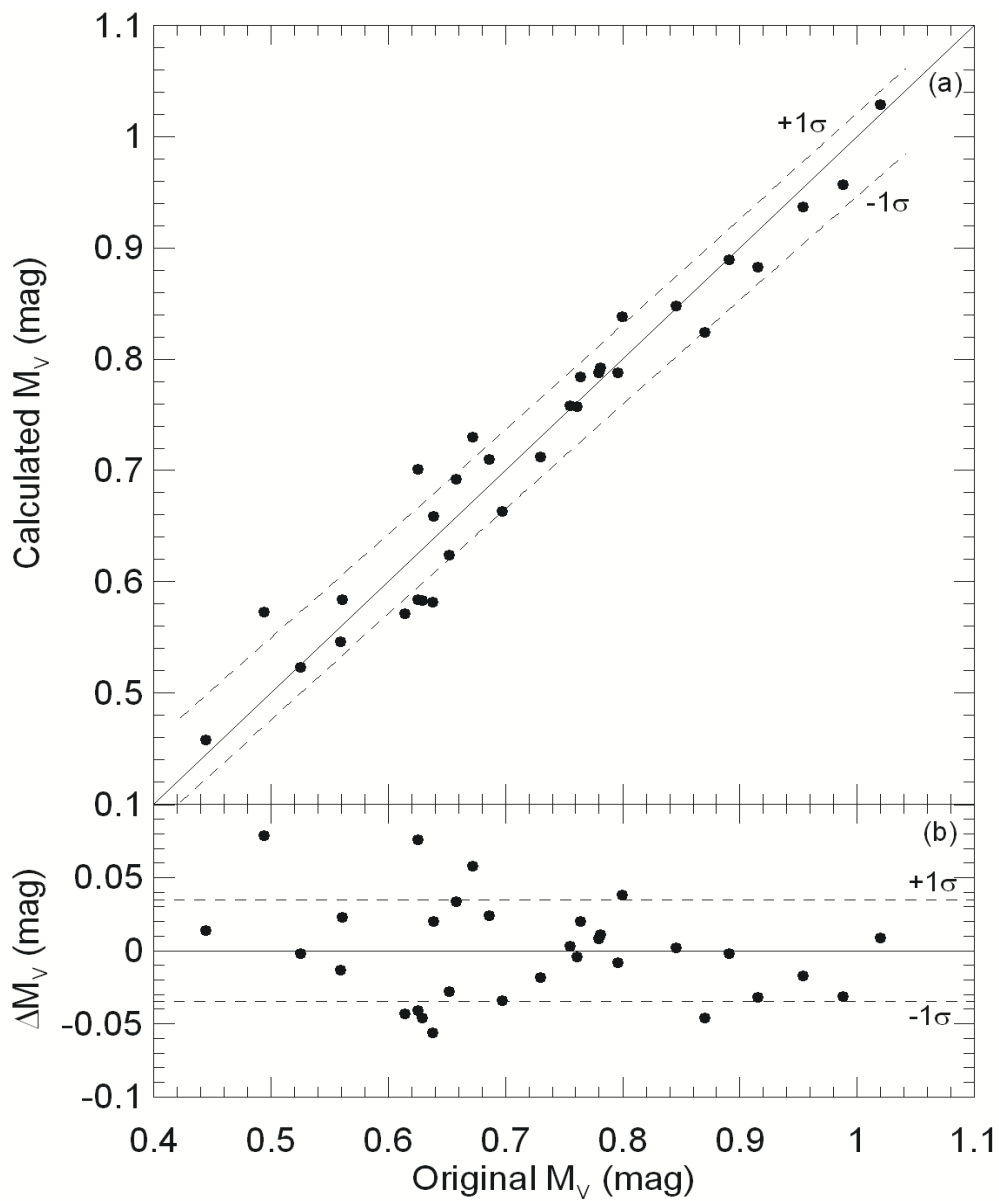}
\caption[]{Comparison of the mean absolute magnitudes of the RC 
stars in Table 1 with those calculated from the 
calibration formula in this study.}
\end{center}
\end{figure}

\begin{figure}
\begin{center}
\includegraphics[scale=0.6, angle=0]{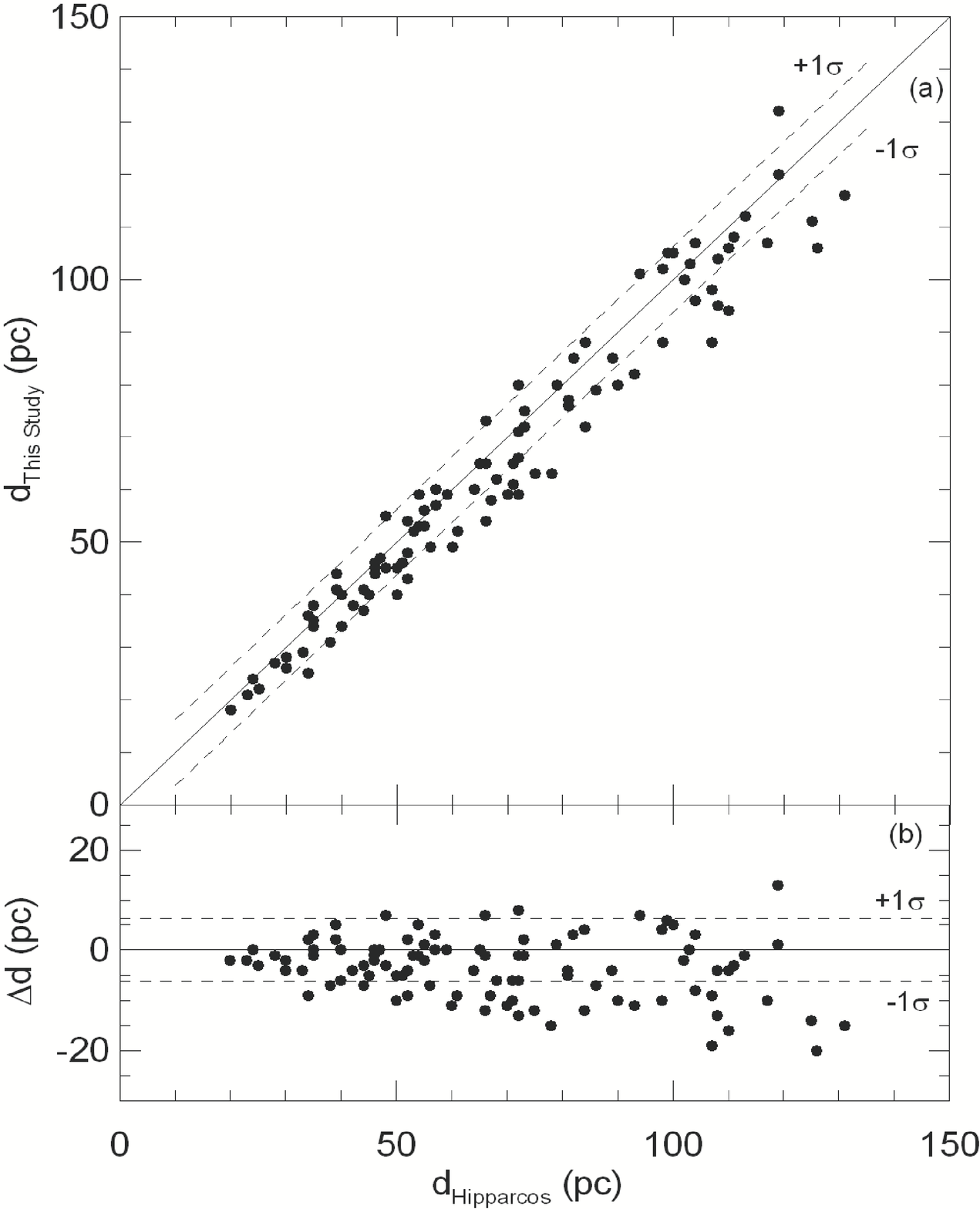}
\caption[]{The comparison of distances estimated using our calibration 
with those taken from \cite{Laney12}.}
\end{center}
\end{figure}

\begin{figure}
\begin{center}
\includegraphics[scale=0.6, angle=0]{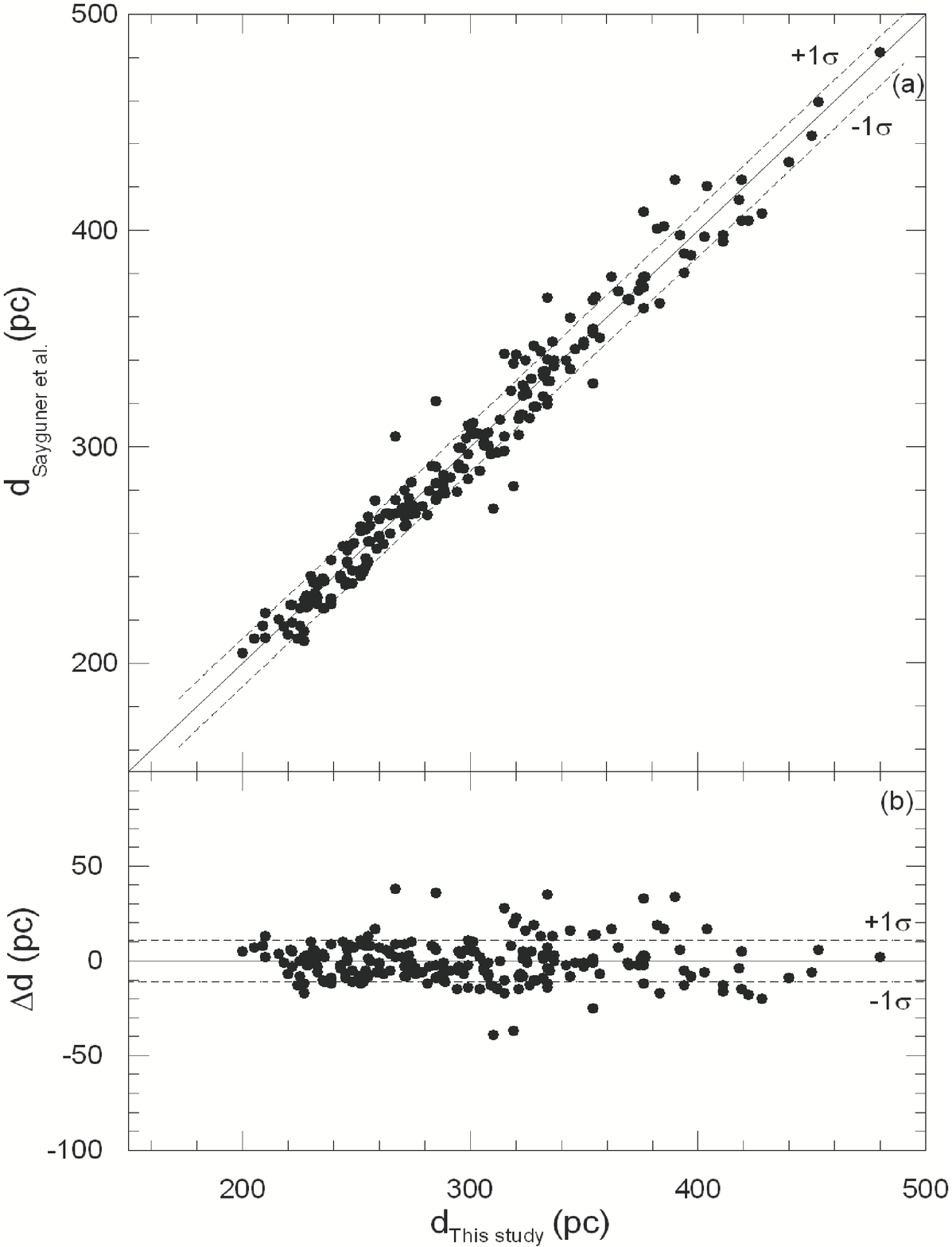}
\caption[] {The comparison of distances estimated using our calibration 
with those taken from \cite{Saguner11}.}
\end{center}
\end{figure}

\end{document}